\documentclass[twocolumns,reprint,superscriptaddress,amsmath,amssymb,prb,nofootinbib]{revtex4-2}

\usepackage{tikz}
\usepackage{import}
\usepackage{graphicx}
\usepackage{amsmath}
\usepackage{amssymb}
\usepackage{amsthm}
\usepackage{mathtools}
\usepackage{physics}
\usepackage{soul}
\usepackage{qcircuit}
\usepackage{subfig}
\usepackage[font=small,labelfont=bf]{caption}

\newtheorem{theorem}{Theorem}[section]

\newtheorem{lemma}{Lemma}[section]

\theoremstyle{remark}
\newtheorem{remark}{Remark}

\usepackage[colorlinks=true,linkcolor=blue,citecolor=blue,urlcolor=blue]{hyperref}

\DeclareMathOperator{\expfun}{exp}  

\DeclareMathOperator{\repart}{\mathfrak{Re}}    
\DeclareMathOperator{\impart}{\mathfrak{Im}}    
\DeclareMathOperator{\hermpart}{\mathrm{Herm}}    
\DeclareMathOperator{\skewhpart}{\mathrm{Skew}}   

\newcommand{\idenm}[1]{\ensuremath{\mathbb{I}_{#1}}}	
\newcommand{\idenmnodim}{\ensuremath{\mathbb{I}}}       
\newcommand{\onesvec}{\ensuremath{\mathbf{1}}}          
\newcommand{\zerosvec}{\ensuremath{\mathbf{0}}}          

\newcommand{\myham}{\ensuremath{\mathcal{H}}}   
\newcommand{\htimedab}{\ensuremath{h}}   

\newcommand{\pauliopx}[1]{\ensuremath{\sigma^{\mathrm{x}}_{#1}}}
\newcommand{\pauliopy}[1]{\ensuremath{\sigma^{\mathrm{y}}_{#1}}}
\newcommand{\pauliopz}[1]{\ensuremath{\sigma^{\mathrm{z}}_{#1}}}
\newcommand{\pauliop}[1]{\ensuremath{\sigma^{\mathrm{#1}}}}
\newcommand{\pauliopi}[2]{\ensuremath{\sigma^{\mathrm{#1}}_{#2}}}

\begin{document}
\title{Quantum Krylov Subspace Diagonalization \\
via Time Reversal Symmetries}

\author{Nicola Mariella}
\email{nicola.mariella@ibm.com}
\affiliation{IBM Quantum, IBM Research Europe, Trinity Business School, Dublin, D02 F6N2 (Ireland)}

\author{Enrique Rico}
\affiliation{EHU Quantum Center and Department of Physical Chemistry, University of the Basque Country UPV/EHU, P.O. Box 644, 48080 Bilbao, Spain}
\affiliation{DIPC - Donostia International Physics Center, Paseo Manuel de Lardizabal 4, 20018 San Sebastián, Spain}
\affiliation{IKERBASQUE, Basque Foundation for Science, Plaza Euskadi 5, 48009 Bilbao, Spain}
\affiliation{European Organization for Nuclear Research (CERN),  Theoretical Physics Department, CH-1211 Geneva, Switzerland}

\author{Adam Byrne}
\affiliation{IBM Quantum, IBM Research Europe, Trinity Business School, Dublin, D02 F6N2 (Ireland)}
\affiliation{School of Mathematics, Trinity College Dublin, Ireland}

\author{Sergiy Zhuk}
\affiliation{IBM Quantum, IBM Research Europe, Trinity Business School, Dublin, D02 F6N2 (Ireland)}
 
\begin{abstract}
Krylov quantum diagonalization methods have emerged as a promising use case for quantum computers. However, many existing implementations rely on controlled operations, which pose challenges to near-term quantum hardware.
We introduce a novel protocol, termed \textit{Krylov Time Reversal} (KTR), that circumvents these bottlenecks by leveraging time-reversal symmetry in Hamiltonian evolution.
Using symmetric time dynamics, we show that it is possible to recover real-valued Krylov matrix elements, which significantly reduces the circuit depth and enhances compatibility with shallow quantum architectures.
Furthermore, the protocol’s structure indirectly reduces the total evolution time, benefiting both near-term and long-term architectures.
We validate our method through numerical simulations on paradigmatic Hamiltonians exhibiting time-reversal symmetry, including the transverse-field Ising model and a lattice gauge theory,
demonstrating accurate spectral estimation and favorable circuit constructions.
\end{abstract}

\maketitle

\section{Introduction}
As quantum devices transition steadily into the pre-fault-tolerant era, there is growing interest in algorithms that can leverage limited coherence times and circuit depths while still extracting useful spectral information from quantum systems. Among the most promising techniques for the estimation of ground state energy in near-term devices are \textit{quantum subspace diagonalization} methods \cite{parrish2019,stair-multiref-quantum-kry,PRXQuantum.2.040352,PRXQuantum.3.020323,PhysRevA.105.022417,PRXQuantum.2.010333,Shen_2023,Kirby_2023,motta-qite,PRXQuantum.3.020323,subspace-methods-review,yoshioka2025krylov,yu2025quantumcentricalgorithmsamplebasedkrylov,doi:10.1126/sciadv.adu9991,duriez2025computingbandgapsperiodic,oumarou2025molecularpropertiesquantumkrylov}. These techniques aim to construct low-dimensional subspaces whose diagonalization yields accurate approximations of the extremal eigenvalues of the full-system Hamiltonian. The most suitable choice of subspace for near-term quantum computers is that generated by powers of the real-time evolution operator, which can be efficiently approximated on existing quantum devices. Methods using this subspace are often referred to as \textit{Krylov Quantum Diagonalization} (KQD) methods \cite{parrish2019,stair-multiref-quantum-kry,PRXQuantum.2.040352,PRXQuantum.3.020323,PhysRevA.105.022417,PRXQuantum.2.010333,Shen_2023}, and possess rigorous convergence guarantees, even in the presence of noise \cite{doi:10.1137/21M145954X,Kirby2024analysisofquantum}.

In many KQD protocols, the construction of Krylov overlap matrices requires controlled unitary operations, which are challenging to implement on near-term quantum devices ~\cite{Preskill_2018}. Recent efforts have focused on developing control-free or control-light versions of KQD that are better suited to the constraints of current quantum hardware. In particular, schemes inspired by the theory of holomorphic functions have demonstrated how spectral properties can be extracted using phase-sensitive measurements with locally controlled operations~\cite{schiffer2025hardwareefficientquantumphaseestimation}. For Hamiltonians with global $U(1)$ or global $SU(2)$ symmetries, multi-fidelity protocols can be used to avoid controlled unitary operations \cite{PhysRevA.105.022417}, and have been experimentally demonstrated on 56 qubits \cite{yoshioka2025krylov}. Other sampling-based KQD methods, which require assumptions on the sparsity of the ground state, use only real-time evolutions and sampling in the computational basis, and have been performed experimentally on 85 qubits \cite{yu2025quantumcentricalgorithmsamplebasedkrylov}. Work in super-Krylov methods \cite{byrne-superkrylov} has also shown how advanced basis construction techniques, tailored to the structure of quantum Hamiltonians, can enhance the precision of subspace approximations while reducing the circuit depth.

Motivated by these developments, we propose a novel KQD algorithm for Hamiltonians with time-reversal symmetry, which avoids the need for controlled time evolutions.
Specifically, the time-symmetric evolution can be exploited to recover real-valued overlap information between the Krylov states. Thus, the method, which we call \textit{Krylov Time Reversal} (KTR), is desirable for implementation in shallow and noisy devices while preserving the favorable convergence properties of KQD. The proposed algorithm builds on the insights of low-control measurement approaches \cite{PhysRevLett.132.220601} and compilation techniques \cite{chowdhury2025-controlization}, as well as the theoretical structure of KQD \cite{parrish2019,stair-multiref-quantum-kry,PRXQuantum.2.040352,PRXQuantum.3.020323,PhysRevA.105.022417,PRXQuantum.2.010333,Shen_2023} and its recent advances \cite{byrne-superkrylov}. Our contribution is a hybrid algorithmic framework that is both resource-efficient and theoretically grounded, offering a path forward for quantum spectral estimation in the near term.

The paper is organized as follows: \autoref{section:quantum-krylov-method} outlines the foundational principles of KQD; \autoref{section:formulation} presents our proposal; \autoref{section:init-state} describes the construction of initial states; and \autoref{section:experiments} evaluates the approach by benchmarking on two representative models.

\section{Krylov Quantum Diagonalization}
\label{section:quantum-krylov-method}
KQD is a variational framework that approximates the low-lying eigenstates of a given Hamiltonian by constructing a \textit{Krylov subspace} \cite{krylov1931,golub} generated by the time evolution applied to some reference state. We introduce the key aspects of KQD that are relevant to our work.

Consider an arbitrary many-body Hamiltonian $\myham{}(\gamma)$ depending on the parameter $\gamma \in \mathbb{R}$. Let $\ket{v_0}$ be a fixed initial state. We define the time evolution (in practice, Trotterized) by
\begin{align}
    \label{eq:v-at-t-def}
	\ket{v(t)}\coloneqq& \expfun\left(-\imath t \myham{}\right) \ket{v_0},
\end{align}
for $t \in \mathbb{R}$, so that $\ket{v(0)}=\ket{v_0}$. Let $t_a, t_b \in \mathbb{R}$ be time displacements. We note the following properties:
\begin{subequations}
\begin{align}
	\label{eq:overlap-vavb}
	\bra{v(t_a)}\ket{v(t_b)} =& \bra{v_0} \expfun\left(-\imath (t_b-t_a)\myham\right) \ket{v_0}\\
	\label{eq:overlap-vavb-toeplitz}
	=& \bra{v_0}\ket{v(t_b-t_a)},\\
	\label{eq:overlap-vahvb}
	\bra{v(t_a)}\myham\ket{v(t_b)} =& \bra{v_0} \expfun\left(\imath t_a\myham\right)\myham{} \ket{v(t_b)}\\
	\label{eq:overlap-vahvb-toeplitz}
	=& \bra{v_0} \myham{} \ket{v(t_b-t_a)}.
\end{align}
\end{subequations}
The identity in \eqref{eq:overlap-vahvb} also holds for any operator $K$ that commutes with the Hamiltonian, that is, $\bra{v(t_a)}K\ket{v(t_b)} = \bra{v_0}K\ket{v(t_b-t_a)}$.

Let $\mathcal{I} \subset \mathbb{R}$ be a set of time displacements of finite cardinality $m$, and consider the pairs $(t_a, t_b) \in \mathcal{I}\times \mathcal{I}$.
The overlaps in \eqref{eq:overlap-vavb} and \eqref{eq:overlap-vahvb} can be used to construct the $m\times m$ matrices\footnote{It is assumed that the dimension $m$ (i.e., the number of Krylov vectors) depends polynomially on the number of qubits. Thus, any further computation on such objects is classical.} $A, B$, whose entries are given by
\begin{subequations}
\begin{align}
	\label{eq:mat_a_entry}
	A_{a, b} =& \bra{v(t_a)}\myham{}\ket{v(t_b)},\\
	\label{eq:mat_b_entry}
	B_{a, b} =& \bra{v(t_a)}\ket{v(t_b)},
\end{align}
\end{subequations}
where natural ordering is assumed for the time displacements. It follows that both matrices $A$ and $B$ are \textit{Hermitian-Toeplitz}\footnote{An $n \times n$ Toeplitz matrix $A$ is defined by the entries $A_{i, j}=a_{i-j}$ depending on the difference $i-j$. If, in addition, $A$ is Hermitian, then a selected row or column defines A.} \cite{Horn_Johnson_2012}. The canonical method is then concluded by solving the \textit{generalized eigenvalue problem} \cite{golub} $A \boldsymbol{x} = \lambda B \boldsymbol{x}$, to obtain the low-lying spectrum of $\myham{}$.

The entries for the matrices in \eqref{eq:mat_a_entry} and \eqref{eq:mat_b_entry} are, in general, in the complex field $\mathbb{C}$, so their estimation requires the \textit{Hadamard test}~\cite{hadamard-test,Nielsen_Chuang_2010}. However, its practical implementation encounters difficulties mainly due to the need for high-fidelity controlled operations, which render this approach unreliable on \emph{Noise-Intermediate-Scale Quantum} (NISQ) devices. In this work, we show that, under certain assumptions on the Hamiltonian $\mathcal{H}(\gamma)$, these overlaps are real-valued and can be obtained as quantum expectations.

\section{Formulation}
\label{section:formulation}
We consider the set of Hamiltonians for which there exists a \textit{Hermitian involutory} operator $T$ such that
\begin{align}
    \label{eq:t-op-anticomm}
    T\myham{}(\gamma)=-\myham{}(\gamma) T,
\end{align}
for all $\gamma \in \mathbb{R}$.
That is, we require the anticommutator relation $\{T, \myham\}=0$, independently of the parameter $\gamma$. In linear algebra, this condition is known as \textit{generalized skew-centrosymmetry} \cite{doi:10.1137/S0895479801386730}.
The definition of $T$ implies unitarity, so for the time evolution we observe that
\begin{align}
    \label{eq:time-evol-reversal}
    Te^{-\imath t\myham{}}T =& e^{-\imath t T\myham{}T} = e^{\imath t \myham{}}.
\end{align}
That is, the conjugate action of $T$ has the effect of \textit{reversing the time evolution}, therefore we call the operator $T$ the \textit{time reversal operator}\footnote{
    This should not be confused with the time reversal in physics, which is obtained from the action of an anti-unitary operator.
}.
If a Hamiltonian $\myham{}$ admits the operator $T$,
then it is not necessarily unique\footnote{We expand on the subject in \autoref{section:time-reversal-symmetry}.}.
However, for the main part of this work, we consider a single (arbitrary) choice of such an operator.

The initial state $\ket{v_0}$ for the time evolution in \eqref{eq:v-at-t-def}, is assumed to be stabilized by the action of $T$ up to a constant sign flip\footnote{The constant $c \in \{\pm 1\}$ is unphysical when interpreted as a global phase; however, it will become relevant in \autoref{section:init-state}.}, that is
\begin{align}
    \label{eq:vzero-stab}
    T \ket{v_0} = c \ket{v_0},
\end{align}
with $c \in \{\pm 1\}$. Moreover, it can be shown that $\ket{v_0}$ consists of a linear combination of eigenvectors of $T$ from the same eigenspace\footnote{
    Since $T$ is Hermitian-involutory, then its eigenvalues are $\{\pm 1\}$, so we have two eigenspaces.
}. The formal construction of the initial state is treated in \autoref{section:init-state}.

The first key result (proof in \autoref{section:proofs}) reveals a convenient formulation for the inner product in \eqref{eq:overlap-vavb}.
\begin{lemma}
    \label{lemma:gram-from-expectation-on-T}
    Consider the Hamiltonian $\myham{}$ and a Hermitian involutory anticommuting operator $T$ fulfilling \eqref{eq:vzero-stab}.
    Let $t_a, t_b \in \mathbb{R}$ and let $\htimedab=\frac{t_b-t_a}{2}$, then
    \begin{align}
        \label{eq:first-key-result}
        \bra{v(t_a)}\ket{v(t_b)} =&
        c\bra{v(\htimedab)}T\ket{v(\htimedab)} \in \mathbb{R},
    \end{align}
    with $c=\bra{v_0}T\ket{v_0} \in \{\pm 1\}$.
\end{lemma}
We observe that the LHS of the expression in \eqref{eq:first-key-result} corresponds to the entry $B_{a, b}$ of the Gram matrix in \eqref{eq:mat_b_entry}. Put differently, this construction yields entries that are guaranteed to be real, which can be obtained without requiring a controlled time evolution. The RHS of \eqref{eq:first-key-result} is an expectation with the Hermitian observable $T$. As an additional benefit, we observe that the time evolution from $t_b-t_a$ in \eqref{eq:overlap-vavb} to $\htimedab=(t_b-t_a)/2$ in \eqref{eq:first-key-result} has halved.

We turn to the second key result (proof in \autoref{section:proofs}).
\begin{lemma}
    \label{lemma:overlap-with-K-from-expectation-on-KT}
    Assuming the same conditions of \autoref{lemma:gram-from-expectation-on-T},
    let $K$ be a Hermitian operator such that $[K, \myham{}]=0$ and $\{K, T\}=0$.
    Then
    \begin{align}
        \label{eq:second-key-result}
    	\bra{v(t_a)}K\ket{v(t_b)} =&
        \imath c \bra{v(\htimedab)} (\imath KT) \ket{v(\htimedab)} \in \imath \mathbb{R},
    \end{align}
    with $c=\bra{v_0}T\ket{v_0} \in \{\pm 1\}$.
\end{lemma}
As a corollary of \autoref{lemma:overlap-with-K-from-expectation-on-KT}, when $K=\myham{}$ we obtain the entries of the matrix $A$ defined in \eqref{eq:mat_a_entry}, that is,
\begin{align}
    \label{eq:mat_a_entry-from-obs-iht}
    A_{a, b} =&
    \imath c \bra{v(\htimedab)} (\imath \myham{}T) \ket{v(\htimedab)} \in \imath \mathbb{R}.
\end{align}
Put differently, the entries of the matrix $A$ are obtained as expectations of the Hermitian observable $\imath \myham{}T$.
The benefit of halved time evolution persists and could also be favorable for \emph{tensor networks} (TN) \cite{Or_s_2014} based computations.

The matrices $A$ and $B$ are Hermitian-Toeplitz, which is reflected by the fact that the state $\ket{v(\htimedab)}$ (which appears in the RHS of \eqref{eq:first-key-result} and \eqref{eq:second-key-result}) depends solely on the difference $t_b-t_a$. Such structures can be defined by their first row\footnote{
    In practice, assuming $t_1=0$, the first row of $B$ is obtained from the expectations $B_{1, j}=c \bra{v(h_j)}T\ket{v(h_j)}$ with $h_j=\frac{t_j-t_1}{2}=\frac{t_j}{2}$ and $t_j \in \mathcal{I}$.
    Then, $B_{i, j}=B_{1, |j-i| + 1}$.
}, so we need to evaluate a linear number of expectations in the number of Krylov vectors.

To conclude, we examine the spectral characterization of the Hamiltonians concerning this work. Given the eigenvalue equation $\myham{}\ket{\psi_i}=\lambda_i \ket{\psi_i}$, we observe that $\myham{}(T\ket{\psi_i})=-\lambda_i (T\ket{\psi_i})$, so $\bra{\psi_i}(T\ket{\psi_i})=0$ when $\lambda_i\ne 0$. That is, the spectrum is mirrored, so the ground and the most excited energies have the same magnitude.

\section{Initial state $\ket{v_0}$}
\label{section:init-state}
We study the general construction for the initial state $\ket{v_0}$ fulfilling \eqref{eq:vzero-stab}.
In addition, we obtain another initial state $\ket{v_0^{\perp}}$, orthogonal to the former, that satisfies the same property.
Let $T$ be a time reversal operator for the Hamiltonian $\myham{}$ and define
\begin{equation}
    \label{eq:projectors-def}
    P \coloneqq \frac{T+\idenmnodim}{2},\quad
    P^{\perp} \coloneqq \idenmnodim-P =\frac{\idenmnodim-T}{2},
\end{equation}
as the (complementary\footnote{
    That is $P^{\perp}=\idenmnodim-P$ with $P^2=P$ and $P^{\dagger}=P$, so $PP^{\perp}=0$.
}) \textit{orthogonal projections} \cite{Horn_Johnson_2012} induced by $T$.
As a result of the direct sum decomposition $V \oplus V^{\perp}$ of the underlying vector space,
any state $\ket{\varphi}$ can be expressed as $\ket{\varphi} = P\ket{\varphi} + P^{\perp}\ket{\varphi}$.
Then, for any state $\ket{\varphi}$ such that $\bra{\varphi}P\ket{\varphi}\ne 0$, we have
$T\ket{v_0} = \ket{v_0}$,
with $\ket{v_0}\coloneqq \xi P\ket{\varphi}$ and $1/\xi^2=\bra{\varphi}P\ket{\varphi}$, $\xi>0$.
In addition, if $\bra{\varphi}P^{\perp}\ket{\varphi}\ne 0$, then
$T\ket{v_0^{\perp}} = -\ket{v_0^{\perp}}$,
with $\ket{v_0^{\perp}}\coloneqq \frac{\xi}{\sqrt{\xi^2-1}} P^{\perp}\ket{\varphi}$.
Moreover, we have the orthogonality $\bra{v_0}\ket{v_0^{\perp}}=0$, when $\ket{v_0}$ and $\ket{v_0^{\perp}}$ are defined. This shows that the constant $c\in \{\pm 1\}$, concerning
\autoref{lemma:gram-from-expectation-on-T} and \autoref{lemma:overlap-with-K-from-expectation-on-KT}, depends on the choice of the projector used for the preparation of $\ket{v_0}$.

In summary, one can apply either the projector $P$ or its complement $P^{\perp}$ on some \emph{arbitrary} state $\ket{\varphi}$ to obtain,
upon normalization (assuming a non-zero projection), the initial state $\ket{v_0}$ that fulfills \eqref{eq:vzero-stab}.

We analyze a practical setup. Consider the operator $T$ and a state $\ket{\varphi}$ that can be efficiently represented as a \textit{matrix product operator} (MPO) \cite{PhysRevB.95.035129} and a \textit{Matrix Product State} (MPS) \cite{Or_s_2014}, respectively. Then, after classically obtaining the projections $P\ket{\varphi}$ and $P^{\perp}\ket{\varphi}$ as MPS states, the corresponding circuits can be prepared using a compiler technique \cite{PhysRevLett.132.040404, 10.1145/3731251}. More generally, the same concept can be extended to initial states represented by TN. A practical circuit construction for projection operators is discussed in \autoref{section:proj-circ}.
In \autoref{section:implicit-Hadamard}, we reveal an indirect connection with the Hadamard test -- the \emph{implicit Hadamard Test}.

\section{Experiments}
\label{section:experiments}
We assess the validity of the method via TN simulations\footnote{The implementation is based on the library \texttt{iTensor} \cite{itensor}. Time evolutions are approximated using the second-order Trotterization \cite{Hatano_2005}, and MPSs have a maximum bond dimension of 500.}, generating Krylov vectors through MPS time evolutions \cite{Paeckel_2019}. We employ as references the canonical KQD\footnote{We note that in the case of KQD, the overlap matrices typically have complex entries, so, its quantum circuit implementation requires the Hadamard test.} and the \textit{density matrix renormalization group} (DMRG)~\cite{SCHOLLWOCK201196}, observing strong agreement; all further details on the setting up of the quantum Krylov subspace diagonalization (e.g., time step choice, spectral thresholding) are addressed in the existing literature \cite{doi:10.1137/21M145954X, parrish2019}. We proceed with the two models that follow.

\subsection{Case: The transverse-field Ising model}
\label{section:example-ising}
We construct an initial state $\ket{v_0}$ for the \textit{Transverse-Field Ising Model} (TFIM), which is guaranteed to have sufficient overlap with the relevant eigenspaces, while reducing the cost of circuit implementation. Recall the definition of the Hamiltonian
\begin{align}
    \label{eq:ising-h0-hd}
    \myham{}(\gamma) =& -\sum_{i} \sigma_i^x \sigma_{i+1}^x
    - \gamma\sum_i \sigma_i^z,
\end{align}
which, for simplicity, is assumed on the Hilbert space of $n=4d$ qubits for some positive integer $d$. The two terms in \eqref{eq:ising-h0-hd}, are denoted as the base terms $\myham{}_0$ and the driving terms $\myham{}_d$, respectively, so $\myham{}=\myham{}_0 + \gamma \myham{}_d$. For the involutory operator $T=(\pauliopy{} \otimes \pauliopx{})^{\otimes (2d)}$, it can be verified that $\{T, \myham{}(\gamma)\}=0$ for all $\gamma \in \mathbb{R}$.

\begin{figure}
    \centering
    \includegraphics[width=0.45\textwidth]{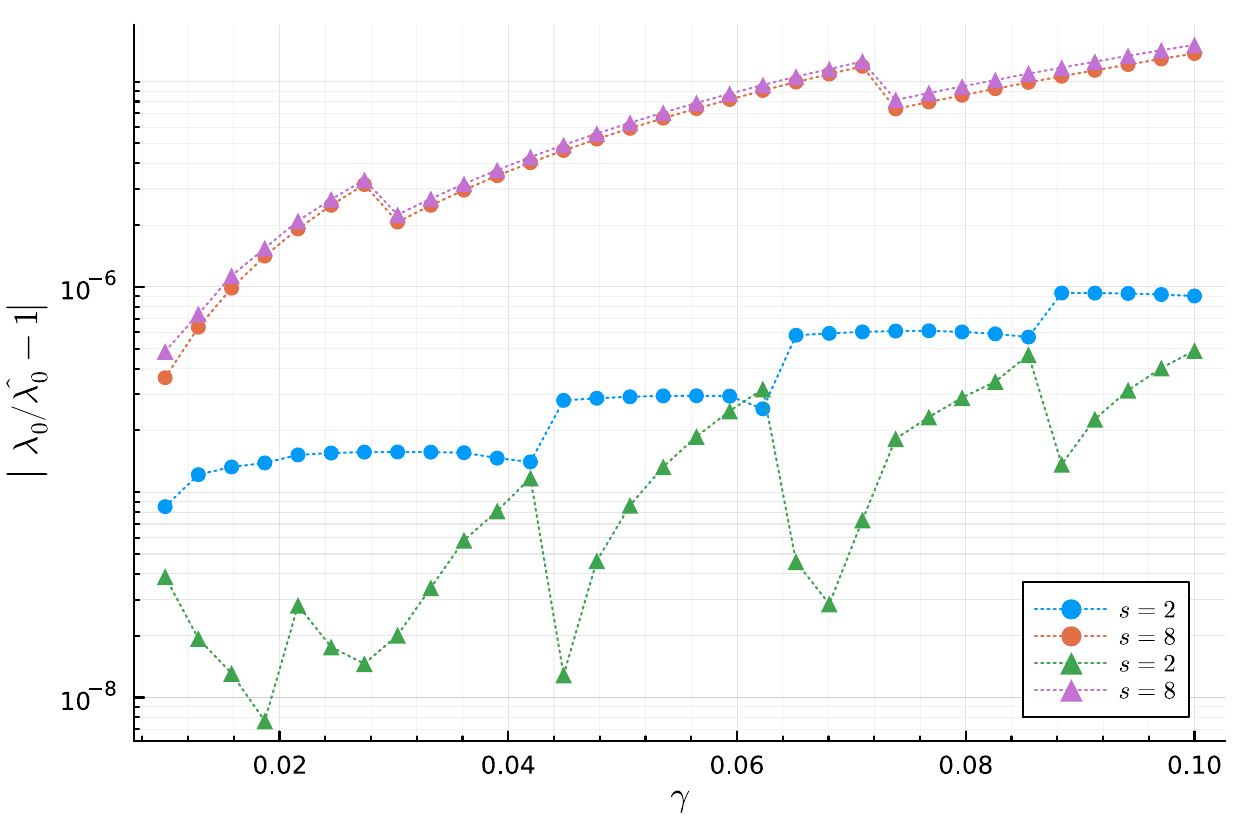}
    \caption{
        Relative error of the ground energy $\lambda_0$ for the TFIM on $n=64$ qubits with $m=128$ (MPS) Krylov vectors and $s$ the number of blocks $\ket{w_0}$.
        References for $\widehat{\lambda}_0$ are KQD ($\bullet$) and DMRG ($\blacktriangle$),
        with initial state $\ket{\varphi}=\ket{+}^{\otimes n}$.
    }
    \label{fig:experiment-tfim}
\end{figure}

Assume that the parameter $|\gamma|$ is small enough for the term $\myham{}_0$ to dominate. Let the state $\ket{w_0}$, on $r=n/s=4d/s$ qubits with $s$ a positive divisor of $d$, be defined by
\begin{align}
    \label{eq:ising-init-state-block-i}
    \ket{w_0} =& \frac{1}{\sqrt{2}} \left(
        (-1)^{r/4} \ket{+}^{\otimes r} + \ket{-+}^{\otimes \frac{r}{2}}
    \right).
\end{align}
\begin{figure}
    \centering
    \begin{equation*}
    \Qcircuit @C=0.7em @R=0.25em {
    \lstick{\ket{1}} & \gate{H} & \ctrl{2} & \gate{H} & \qw\\
    \lstick{\ket{0}} & \gate{H} & \qw      & \qw      & \qw\\
    \lstick{\ket{0}} & \qw      & \targ    & \gate{H} & \qw\\
    \lstick{\ket{0}} & \gate{H} & \qw      & \qw      & \qw
    }
    \end{equation*}
    \caption{Block circuit (case $r=4$ qubits) for the preparation of the state $-\ket{w_0}$ in \eqref{eq:ising-init-state-block-i}.
    The product $\ket{v_0}=\ket{w_0}^{\otimes s}$ yields the initial state.
    }
    \label{fig:ising-init-state-block-i}
\end{figure}
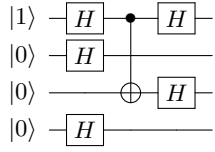
We obtain an initial state $\ket{v_0}=\ket{w_0}^{\otimes s}$ fulfilling \eqref{eq:vzero-stab}, as the $s$-fold product of blocks $\ket{w_0}$.
We justify this construction for its overlap with the ground state (possibly degenerate) of the base term $\myham{}_0$, which is dominant for small $|\gamma|$. In fact, we can verify that $|\bra{+}^{\otimes n}\ket{v_0}| = 1/\sqrt{2^s}$.
So, for relatively small divisors $s$ of $d$, we guarantee the required symmetry and overlap with the ground state.
Another rationale is to avoid generating long-range entanglement throughout the system \cite{javadiabhari2025bigcatsentanglement120},
although methods based on dynamic circuits are available \cite{PRXQuantum.5.030339}.
A formal treatment of this kind of structure is presented in \autoref{section:proj-circ}. The experimental results for this construction are reported in \autoref{fig:experiment-tfim}.

The expression for the Hermitian observable $\imath \myham{}T$, determining the expectation in \eqref{eq:mat_a_entry-from-obs-iht},
follows directly from the identity $\imath \pauliopx{} \pauliopy{} \pauliopz{} =-\idenmnodim$, that is,
\begin{align}
    \imath \myham{}T
    =& 
    \sum_{i} (\sigma_i^z\sigma_i^y) \sigma_{i+1}^x T
    +\gamma\sum_i (\sigma_i^y\sigma_i^x) T.
\end{align}

\subsection{Case: $\mathbb{Z}_2$ gauge Higgs model}
\label{section:example-lgt}
\begin{figure}
    \centering
    \includegraphics[width=0.45\textwidth]{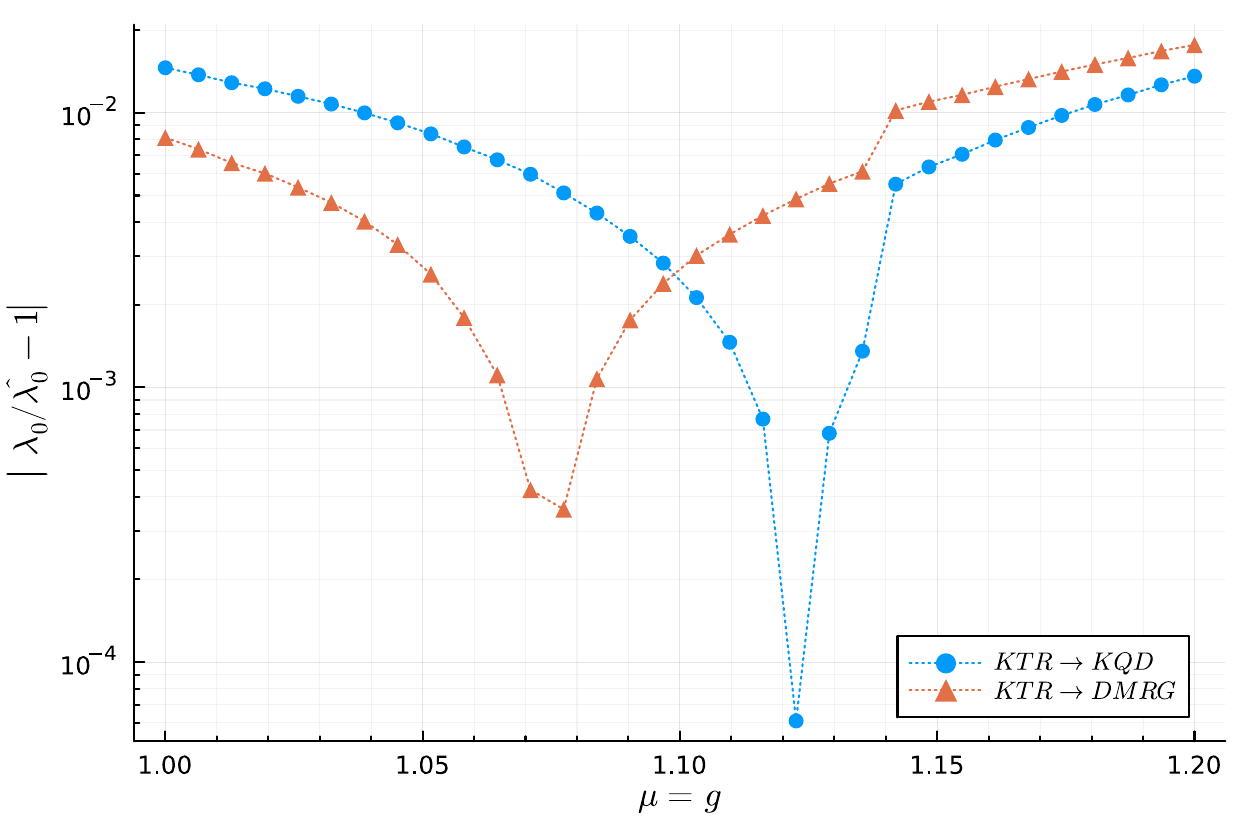}
    \caption{
        Relative error for the $\mathbb{Z}_2$ gauge Higgs model on $n=64$ qubits, relative to the ground energy for the sector $G\ket{\psi}=\ket{\psi}$.
        We considered $m=80$ Krylov vectors (MPS) and $s=2$ blocks $\ket{w_0}$.
    }
    \label{fig:experiment-lgt}
\end{figure}

We benchmark a model from high-energy physics: a \textit{lattice gauge theory} (LGT) Hamiltonian, specifically the \emph{$\mathbb{Z}_2$ gauge Higgs model}~\cite{cobos2025, PhysRevD.19.3682, PhysRevX.15.011001}.
Such models capture essential features of gauge-invariant dynamics on a discretized space, providing a rich testbed for exploring quantum many-body phenomena.
Denoting, respectively, $\mathcal{V}$ and $\mathcal{L}$, the sets of vertices and links of the lattice, the Hamiltonian is defined by
\begin{align}
    \label{eq:lgt}
    \myham{}(\mu, g) =& -\sum_{i  \in \mathcal{L}} \pauliopi{z}{i-1}\pauliopi{z}{i}\pauliopi{z}{i+1}
    - \mu \sum_{j \in \mathcal{V}} \pauliopi{x}{j}
    - g \sum_{k \in \mathcal{L}} \pauliopi{x}{k},
\end{align}
where the interaction term corresponds to the minimal coupling of vertices (matter) qubits and links (gauge) qubits; the onsite term on the vertices characterizes the energy cost to excite a matter qubit, while the onsite term on the links describes the energy cost to excite a gauge qubit. A time reversal operator is $T=(\pauliopy{})^{\otimes n}$.
The gauge symmetries are generated by the operators $G_k$ \cite{cobos2025} and let $G$ be their averaged sum, which can be verified to anticommute with $T$ and commute with $\myham{}$.
Also, $G$ has spectral norm $1$.
In relation to \autoref{section:init-state}, we start with the state $\ket{\varphi}=\ket{+}^{\otimes n}$, which is gauge invariant since $G\ket{\varphi}=\ket{\varphi}$.
We aim at obtaining the ground energy for the sector (invariant subspace) determined by the latter.
The initial state is $\ket{v_0}=\xi P \ket{\varphi}=\xi (\ket{\varphi} + T\ket{\varphi})/2$, but $G(T\ket{\varphi})=-T\ket{\varphi}$, so $\ket{v_0}$ is a superposition from two sectors.
However, we implement a penalty for the complementary sector using \autoref{lemma:overlap-with-K-from-expectation-on-KT} with $K=\eta (\idenmnodim - G)$, for some tradeoff parameter $\eta > 0$.
Finally, as in \autoref{section:example-ising}, we break down the initial state $\ket{v_0}$ into an $s$-fold product of blocks $\ket{w_0}$.
The experimental results for this construction are reported in \autoref{fig:experiment-lgt}.

\clearpage
\section*{Acknowledgments}
The data generated from quantum simulations and supporting this study are available from the authors upon reasonable request.

N.M. is grateful to Antonio Mezzacapo (IBM Research),
Albert Akhriev (IBM Research),
Dmytro Mishagli (IBM Research),
Jes\'us Cobos (UPV/EHU), William Kirby (IBM Research), and Pawel Wocjan (IBM Research) for useful technical discussions.

E.R. acknowledges the financial support received from the IKUR Strategy under the collaboration agreement between the Ikerbasque Foundation and UPV/EHU on behalf of the Department of Education of the Basque Government. E.R. acknowledges support from the BasQ strategy of the Department of Science, Universities, and Innovation of the Basque Government. E.R. is supported by the grant PID2021-126273NB-I00 funded by MCIN/AEI/10.13039/501100011033 and by “ERDF A way of making Europe” and the Basque Government through Grant No. IT1470-22. This work was supported by the EU via QuantERA project T-NiSQ grant PCI2022-132984 funded by MCIN/AEI/10.13039/501100011033 and by the European Union “NextGenerationEU”/PRTR. This work has been financially supported by the Ministry of Economic Affairs and Digital Transformation of the Spanish Government through the QUANTUM ENIA project, called Quantum Spain project, and by the European Union through the Recovery, Transformation, and Resilience Plan – NextGenerationEU within the framework of the Digital Spain 2026 Agenda.

This work has been partially funded by the Eric \& Wendy Schmidt Fund for Strategic Innovation through the CERN Next Generation Triggers project under grant agreement number SIF-2023-004.

\clearpage

\appendix
\onecolumngrid

\section{Proofs not included in the main text}
We present the proofs of the main results. We recall that $T$ is a Hermitian involutory operator such that $\{\myham{}, T\}=0$, and that the initial state $\ket{v_0}$ fulfills the property of $T\ket{v_0}=c\ket{v_0}$ with $c\in \{\pm 1\}$. For time displacements $t_a, t_b \in \mathbb{R}$, the half difference is $\htimedab=(t_b-t_a)/2$.
\label{section:proofs}
\begin{proof}[Proof of \autoref{lemma:gram-from-expectation-on-T}]
    \begin{subequations}
    \begin{align}
        \bra{v(t_a)}\ket{v(t_b)} \underset{\eqref{eq:overlap-vavb}}{=}& \bra{v_0}\ket{v(t_b - t_a)}\\
        \underset{\eqref{eq:v-at-t-def}}{=}& \bra{v_0} e^{-\imath \htimedab{} \myham{}} \ket{v(\htimedab)}\\
        \underset{\eqref{eq:time-evol-reversal}}{=}& \bra{v_0} Te^{\imath \htimedab{} \myham{}} T \ket{v(\htimedab{})}\\
        \underset{\eqref{eq:vzero-stab}}{=}& c\bra{v_0} e^{\imath \htimedab{} \myham{}} T \ket{v(\htimedab{})}\\
        =& c\bra{v(\htimedab)}T\ket{v(\htimedab)}.
    \end{align}
    \end{subequations}
\end{proof}

\begin{proof}[Proof of \autoref{lemma:overlap-with-K-from-expectation-on-KT}]
    We proceed in such a way that the inner product becomes a quadratic form, so
    \begin{subequations}
    \begin{align}
    	\bra{v(t_a)}K\ket{v(t_b)} \underset{\eqref{eq:overlap-vahvb}}{=}& \bra{v_0}K\ket{v(t_b-t_a)}\\
        =& \bra{v_0} K e^{-\imath \htimedab{} \myham{}} \ket{v(\htimedab)}\\
        \underset{\eqref{eq:time-evol-reversal}}{=}& \bra{v_0} K Te^{\imath \htimedab{} \myham{}}T \ket{v(\htimedab{})}\\
        =& -\bra{v_0} T K e^{\imath \htimedab{} \myham{}}T \ket{v(\htimedab)}\\
        \underset{\eqref{eq:vzero-stab}}{=}& -c \bra{v_0} K e^{\imath \htimedab{} \myham{}}T \ket{v(\htimedab)}\\
        =& -c \bra{v_0} e^{\imath \htimedab{} \myham{}} (KT) \ket{v(\htimedab)}\\
        =& \imath c \bra{v(\htimedab)} (\imath KT) \ket{v(\htimedab)}.
    \end{align}
    \end{subequations}
    Finally, since the Hermitian operators $K$ and $T$ anti-commute, then $(\imath KT)^{\dagger}=-\imath TK = \imath KT$, hence $\imath KT$ is Hermitian.
\end{proof}

\section{Time reversal symmetry}
\label{section:time-reversal-symmetry}
We provide a construction for Hamiltonians that fulfills the time-reversal symmetry. Let $\mathcal{G}_n=\{\idenmnodim, \pauliopx{}, \pauliopy{}, \pauliopz{}\}^{\otimes n}$ be the set of $n$-qubit Pauli strings. For the remainder of this section, we consider the class of Hamiltonians defined by a linear combination of Pauli strings, that is,
\begin{align}
    \label{eq:pauli-str-ham-for-time-rev}
    \myham{} =& \sum_i h_i \boldsymbol{\sigma}_i,
\end{align}
with coefficients $h_i \in \mathbb{R}$. The result that follows is a generalization of the LGT case in \autoref{section:example-lgt}.
\begin{lemma}
    \label{lemma:ham-odd-iteractions}
    Let $\{\pauliop{a}, \pauliop{b}, \pauliop{c}\}$ be the image of any bijection on the set of Paulis.
    Consider the case of the Hamiltonian in \eqref{eq:pauli-str-ham-for-time-rev}, when each term consists of an odd number of Paulis (other than the identity)
    from the set $\{\pauliop{a}, \pauliop{c}\}$\footnote{This means that each term can also include the (tensor) factors $\{\pauliop{b}, \idenmnodim\}$, but without any parity requirements.}.
    Let $T=\Pi_{i=1}^n \pauliopi{b}{i}$ (so $T=T^{\dagger}$ and $T^2=\idenmnodim$), then we have $T\myham{}T=-\myham{}$, as required in \eqref{eq:t-op-anticomm}.
\end{lemma}
We remark that the preceding lemma can be extended further by considering a family of mutually anticommuting Hermitian matrices instead of the set of Paulis. The proof will naturally follow from the next result.

We outline a general method for obtaining involutory Hermitian anticommuting operators by adapting the procedure proposed in \cite{bravyi2017taperingqubitssimulatefermionic}[Section VIII], intended to identify the generators of Hermitian operators that commute with the Hamiltonian. We parameterize each term $\boldsymbol{\sigma}_i$ of $\myham{}$ using a binary vector $(\begin{array}{c|c} f_x & f_z \end{array})^{\top} \in \{0, 1\}^{2n}$, where the bits of each component $f_x, f_z$ are set according to the Pauli string configuration. For example $\pauliopx{} \otimes \pauliopy{} \otimes \pauliopz{} \otimes \idenmnodim \mapsto (\begin{array}{cccc|cccc}1 & 1 & 0 & 0 & 0 & 1 & 1 & 0\end{array})^{\top}$. The parameterizations of all terms of $\myham{}$ (with $h_i\ne 0$) are collected as the rows of the parity matrix,
\begin{align}
    F =& \left(
    \begin{array}{c|c}
        F_x &  F_z
    \end{array}
    \right).
\end{align}
We solve the linear system $F \mathbf{t}=\onesvec{}$ over the field $\mathbb{F}_2$ for $\mathbf{t}=(\begin{array}{c|c} t_z & t_x \end{array})^{\top}$, where $\onesvec{}$ is the column vector of ones. When solutions exist, the binary vector $\mathbf{t}$ represents the anticommuting and involutory Hermitian operator $T$ for the time reversal. We note that the components of $\mathbf{t}$ (that is, $t_z, t_x$), appear in reversed order w.r.t. the columns of matrix $F$ (i.e. $F_x, F_z$). This construction mimics the interaction of non-commuting Pauli operators, so the equation\footnote{Here $\mathbf{f} \cdot \mathbf{t}$ denotes the inner product of $\mathbf{f}, \mathbf{t}$ on $\mathbb{F}_2$.} $(\begin{array}{c|c} f_x & f_z \end{array})^{\top} \cdot (\begin{array}{c|c} t_z & t_x \end{array})^{\top} =1$, over $\mathbb{F}_2$, holds when the represented operators anti-commute.

The linear system stated earlier is an instance of the \textit{XOR satisfiability problem} (XORSAT) -- a special case of the \textit{boolean satisfiability problem} (SAT) \cite{knuth2021art}. Unlike the general SAT problem, which is NP-complete, XORSAT admits a polynomial-time solution via Gaussian elimination.

We demonstrate the above method applied to the \textit{cluster Hamiltonian} \cite{PhysRevResearch.4.L022020} on a spin-$1/2$ chain. Let $L$ be the $(n-1) \times n$ matrix defined by $L_{i, j}=\delta_{0 \leq (j-i) \le 1}$, and denote by $\idenm{n}$ the $n\times n$ identity matrix. From the definition of the Hamiltonian 
\begin{align}
    \label{eq:cluster-ham-def}
    \myham{} =&
    -g_x \sum_{i=1}^n \pauliopi{x}{i}
    -g_{zz} \sum_i \pauliopi{z}{i}\pauliopi{z}{i+1}
    +g_{zxz} \sum_i \pauliopi{z}{i}\pauliopi{x}{i+1}\pauliopi{z}{i+2},
\end{align}
we construct the $(3n-3) \times (2n+1)$ augmented matrix for linear system $F \mathbf{t}=\onesvec{}$, that is,
\begin{align}
    \left(
    \begin{array}{c|c|c}
        F_x &  F_z & \onesvec{}
    \end{array}
    \right)
    =&
    \left(
    \begin{array}{c|c|c}
        \idenm{n} & 0 & \onesvec{} \\
        0 & L & \onesvec{} \\
        G_x & G_z & \onesvec{}
    \end{array}
    \right),
\end{align}
where $(G_x\,|\,G_z)$ describes the $n-2$ terms of the form $\pauliopi{z}{i}\pauliopi{x}{i+1}\pauliopi{z}{i+2}$. However, it can be shown that the last block row is a linear combination (on $\mathbb{F}_2$) of the first two blocks, so the system is taken to the \textit{reduced row echelon form} (RREF) from this intermediate step, 
\begin{align}
    \label{eq:rref-cluster-intermediate}
    \begin{pmatrix}
        \idenm{n} & \\
        & R
    \end{pmatrix}
    \left(
    \begin{array}{c|c|c}
        \idenm{n} & 0 & \onesvec{} \\
        0 & L & \onesvec{}
    \end{array}
    \right) =&
    \left(
    \begin{array}{c|cc|c}
        \idenm{n} & 0 & \zerosvec{} & \onesvec{}\\
        0 & \idenm{n-1} & \onesvec{} & (R \onesvec{})
    \end{array}
    \right),
\end{align}
where $R$ is the $(n-1) \times (n-1)$ upper-triangular matrix of ones (representing elementary row operations).
The vector $R\onesvec{}$ of row sums of the matrix $R$, consists of alternating zeros and ones, that is $(R\onesvec{})_i=(1-(-1)^{n-i})/2$, for $i=1, 2, \ldots, n-1$. Now, note that the $(2n-1)$-th row vector (i.e., last) of the RHS matrix in \eqref{eq:rref-cluster-intermediate} is always of the form $(0\;0\ldots 0\;1\;1\;|\;1)$. Then, assuming $n$ is even, by backward substitution, we obtain the solutions
$T_1=(\pauliopy{}\otimes \pauliopz{})^{\otimes \frac{n}{2}}$ and $T_2=(\pauliopz{}\otimes \pauliopy{})^{\otimes \frac{n}{2}}$.

As a negative example, we show that for the general \textit{quantum Heisenberg model} given by the Hamiltonian\footnote{For simplicity, we omit the terms related to the external field.}
\begin{align}
    \myham{} =& -\frac{1}{2} \sum_{j=1}^{n-1} \left(J_x \pauliopi{x}{j}\pauliopi{x}{j+1}
    + J_y \pauliopi{y}{j}\pauliopi{y}{j+1}
    + J_z \pauliopi{z}{j}\pauliopi{z}{j+1}
    \right),
\end{align}
the system $F \mathbf{t}=\onesvec{}$ admits no solutions.
Here, the symbols $J_x, J_y, J_z \in \mathbb{R}$ denote the (arbitrary) coupling constants. The $3(n-1) \times (2n + 1)$ augmented matrix for the linear system is
\begin{align}
    \left(
    \begin{array}{c|c|c}
        F_x &  F_z & \onesvec{}
    \end{array}
    \right) =&
    \left(
    \begin{array}{c|c|c}
        L & 0 & \onesvec{} \\
        0 & L & \onesvec{} \\
        L & L & \onesvec{}
    \end{array}
    \right).
\end{align}
In the attempt to obtain the RREF, by adding over $\mathbb{F}_2$ the first two blocks of rows to the third one, we get the block row $\left(\begin{array}{c|c|c}0 & 0 & \onesvec{}\end{array}\right)$, which is inconsistent,
so this Hamiltonian does not admit any time reversal operator.

\section{Implicit Hadamard test}
\label{section:implicit-Hadamard}
We establish the \textbf{Hadamard test} as a linear combination of expectations. We note that this scheme can also benefit other quantum methods that require the Hadamard test,
beyond the present application on Krylov subspace methods.
In addition, we show that the symmetry for the initial state $\ket{v_0}$ stated in \eqref{eq:vzero-stab} can be implicitly relaxed. First, we remind the structure of the Hadamard test and its relation to our work. As considered in \autoref{section:init-state}, let $\ket{\varphi}$ be an arbitrary initial state\footnote{
    In this section we follow the conventions of \autoref{section:init-state}. So, the state $\ket{\varphi}$ is arbitrary in the sense that
    it is not stabilized up to a sign flip by the time reversal operator $T$ (i.e. condition \eqref{eq:vzero-stab}).
    On the other hand, the initial states $\ket{v_0}$ and $\ket{v_0^{\perp}}$ are the result of the projections $P$ and $P^{\perp}$ (defined in \eqref{eq:projectors-def}) on $\ket{\varphi}$, respectively.
} (i.e., not necessarily fulfilling \eqref{eq:vzero-stab}) and let $U=\exp(-\imath (t_b-t_a)\myham{})$ be the unitary corresponding to the time evolution for the time displacement $t_b-t_a$. The Hadamard test consists of the following circuit structure,
\begin{equation}
    \label{eq:hadamard-test-circuit}
    \Qcircuit @C=0.7em @R=0.5em {
        \lstick{\ket{0}}    & \qw     & \gate{H} & \ctrl{1} & \gate{S^p} & \gate{H} & \meter & \cw \\
        \lstick{\ket{\varphi}}  & \qw {/} & \qw      & \gate{U} & \qw & \qw      & \qw
    }
\end{equation}
which is used to estimate separately, the real and imaginary parts of $\bra{\varphi}U\ket{\varphi}$, given by the identity
\begin{align}
    \label{eq:hadamard-test-re-im-parts}
    (1-p)\repart \bra{\varphi}U\ket{\varphi} + p \impart \bra{\varphi}U\ket{\varphi} = \langle \pauliopz{} \rangle_p,
\end{align}
with $p \in \{0, 1\}$, and $\langle \cdot \rangle_p$ the expectation of the given operator w.r.t. the state prepared by the circuit in \eqref{eq:hadamard-test-circuit}.
We note that the quantity $\bra{\varphi}U\ket{\varphi}$ corresponds to the entry $(a, b)$ of the Gram matrix in \eqref{eq:mat_b_entry}, when the initial state $\ket{v_0}$ is an arbitrary state
(i.e., without the stabilizer property in \eqref{eq:vzero-stab}).

We introduce some notational details.
For any complex square matrix $A$, we define the Hermitian and skew-Hermitian parts respectively by
\begin{align}
    \hermpart(A)\coloneqq\frac{A+A^{\dagger}}{2},\quad&\skewhpart(A)\coloneqq\frac{A-A^{\dagger}}{2},
\end{align}
thus $\hermpart(A)+\skewhpart(A)=A$.
When the argument of $\hermpart(\cdot)$ and $\skewhpart(\cdot)$ is an operator time evolution, that is $U=\expfun\left(-\imath t \myham{}\right)$, then we have
the special case
\begin{subequations}
\begin{align}
    \label{eq:skew-and-herm-parts-to-sin-cos}
    \hermpart(U) =& \cosh(-\imath t \myham{}) = \cos(t\myham{}),\\
    \skewhpart(U) =& \sinh(-\imath t \myham{}) = -\imath \sin(t\myham{}),
\end{align}
\end{subequations}
which is relevant for the theory that follows.
Before proceeding further, we obtain some useful identities.
These equations relate the Hamiltonian $\myham{}$ and the action of the projectors in \eqref{eq:projectors-def} on the time evolution.
\begin{lemma}
    \label{lemma:proj-identities}
    Consider the projectors in \eqref{eq:projectors-def}, induced by the time reversal operator $T$, then
    \begin{subequations}
    \begin{align}
        \label{eq:ident-proj-hamilt}
        P\myham{} = \myham{} P^{\perp},\quad&
        P^{\perp}\myham{} = \myham{} P,\\
        \label{eq:ident-proj-skewhp-time-evol}
        P^{\perp} U P = \skewhpart(U)P,\quad&
        P U P^{\perp} = \skewhpart(U) P^{\perp},\\
        \label{eq:ident-proj-hermp-time-evol}
        P U P = \hermpart(U) P,\quad&
        P^{\perp} U P^{\perp} = \hermpart(U) P^{\perp},\\
        \label{eq:ident-pup-with-sqrt-u-and-time-rev-op}
        PUP = P\sqrt{U}^{\dagger}T\sqrt{U}P,\quad&
        P^{\perp} U P^{\perp} = -P^{\perp} \sqrt{U}^{\dagger}T\sqrt{U} P^{\perp},
    \end{align}
    \end{subequations}
    with $U=\expfun\left(-\imath t \myham{}\right)$ and $\myham{}$ the Hamiltonian.
\end{lemma}
\begin{proof}
    The first set of equalities in \eqref{eq:ident-proj-hamilt} follows directly from the anticommutator relation $\{\myham{}, T\}=0$.
    We prove the leftmost in \eqref{eq:ident-proj-skewhp-time-evol}, then the other on the same line, and those in \eqref{eq:ident-proj-hermp-time-evol} follow similarly.
    So
    \begin{subequations}
    \begin{align}
        P^{\perp} U P \underset{\eqref{eq:projectors-def}}{=}&
        \frac{1}{4}\left(
            U+UT-TU-TUT
        \right)\\
        \underset{\eqref{eq:time-evol-reversal}}{=}& 
        \frac{1}{2}\left(
            \frac{(U-U^{\dagger})}{2}+\frac{(U-U^{\dagger})}{2}T
        \right)\\
        =& \skewhpart(U)P.
    \end{align}
    \end{subequations}

    The identities in \eqref{eq:ident-pup-with-sqrt-u-and-time-rev-op} can be deduced directly from the proof of \autoref{lemma:gram-from-expectation-on-T}
    and the definition of initial states in \autoref{section:init-state}.
    Moreover, the $\sqrt{U}$ corresponds to the half-time evolution with $h=t/2$, and the negative sign in the rightmost equality corresponds to the case $c=-1$
    in \autoref{lemma:gram-from-expectation-on-T}.
\end{proof}

\begin{theorem}
    \label{thm:gram-hadamard-test-implicit}
    Let $P, P^{\perp}$ be the projections defined in \eqref{eq:projectors-def}.
    Consider some state $\ket{\varphi}$ assumed\footnote{
        In addition, we remark that the state $\ket{\varphi}$ is not required to fulfill the symmetry in
        \eqref{eq:vzero-stab}.
        The simpler case in which the state $\ket{\varphi}$ belongs to the null space of either $P$ or $P^{\perp}$, is omitted.
    } not to be in the null space of the projections $P, P^{\perp}$.
    Let
    \begin{subequations}
    \begin{align}
        \label{eq:gram-hadamard-test-implicit-first-state}
    	\ket{v(t)}=& \xi \expfun\left(-\imath t \myham{}\right) P\ket{\varphi},\\
    	\ket{v^{\perp}(t)}=& \frac{\xi}{\sqrt{\xi^2-1}} \expfun\left(-\imath t \myham{}\right) P^{\perp}\ket{\varphi},
    \end{align}
    \end{subequations}
    for some normalization coefficient $\xi > 0$ (independent of time $t$), such that $\bra{v(t)}\ket{v(t)}=\bra{v^{\perp}(t)}\ket{v^{\perp}(t)}=1$.
    Then for any pair of time displacements $t_a, t_b \in \mathbb{R}$,
    \begin{equation}
        \label{eq:gram-hadamard-test-implicit}
        \boxed{
        \frac{1}{\xi^2}\bra{v(\htimedab)}T\ket{v(\htimedab)} -
        \left(1-\frac{1}{\xi^2}\right) \bra{v^{\perp}(\htimedab)}T\ket{v^{\perp}(\htimedab)} =
        \repart \bra{\varphi}U\ket{\varphi},}
    \end{equation}
    with $U=\exp\left(-\imath (t_b-t_a)\myham{}\right)$, $\htimedab{}=(t_b-t_a)/2$, $1/\xi^2=\bra{\varphi}P\ket{\varphi}$ and $T$ the Hermitian-involutory operator defining the projections $P, P^{\perp}$.
\end{theorem}
As anticipated, the RHS of the claim in \eqref{eq:gram-hadamard-test-implicit} corresponds to the quantity estimated using the Hadamard test in \eqref{eq:hadamard-test-re-im-parts}
with $p=0$ (i.e., the real part).
The imaginary part can be estimated, up to a sign (i.e., possibly complex conjugate), using the identity
\begin{align}
    |\bra{\varphi}U\ket{\varphi}|^2 =
    \left(\repart \bra{\varphi}U\ket{\varphi}\right)^2 +
    \left(\impart \bra{\varphi}U\ket{\varphi}\right)^2
\end{align}
where the LHS is measured from the usual compute-uncompute circuit.

\begin{proof}
    For a complex scalar $z\in \mathbb{C}$ we denote by $\repart z\coloneqq(z+\overline{z})/2$ its real part.
    First, we note that
    \begin{align}
        \label{eq:re-innerp-equiv-herm-part}
        \repart \bra{\varphi}U\ket{\varphi}=\bra{\varphi}\hermpart(U)\ket{\varphi}.
    \end{align} 
    So, we aim to prove the following
    \begin{align}
        \label{eq:gram-hadamard-test-implicit-proof-i}
        \frac{1}{\xi^2} \bra{v(t_a)}\ket{v(t_b)} +
        \left(1-\frac{1}{\xi^2}\right) \bra{v^{\perp}(t_a)}\ket{v^{\perp}(t_b)} =&
        \bra{\varphi}\hermpart(U)\ket{\varphi},
    \end{align}
    which, as a result of \autoref{lemma:gram-from-expectation-on-T} (applied to the LHS), is equivalent to the claim\footnote{
        We note that the minus sign on the second term in \eqref{eq:gram-hadamard-test-implicit}, arises from $c=\bra{v_0^{\perp}}T\ket{v_0^{\perp}}=-1$,
        since the initial state $\ket{v_0^{\perp}}$ is prepared using the projector $P^{\perp}$.
        See \autoref{section:init-state} for further details.
    } in \eqref{eq:gram-hadamard-test-implicit}.

    By the complementarity of the projectors in \eqref{eq:projectors-def}, we have that $P+P^{\perp}=\idenmnodim$,
    so by summing up the identities in \eqref{eq:ident-proj-hermp-time-evol} we obtain
    \begin{align}
        \label{eq:sum-pup-ptupt}
        P U P + P^{\perp} U P^{\perp} = \hermpart(U) (P + P^{\perp}) = \hermpart(U).
    \end{align}
    We recall that projectors $P, P^{\perp}$ are used in the construction of the initial vectors $\ket{v_0}, \ket{v_0^{\perp}}$
    as outlined in \autoref{section:init-state}.
    So, for example, we can rewrite the state in \eqref{eq:gram-hadamard-test-implicit-first-state} as $\ket{v(t)}= \expfun\left(-\imath t \myham{}\right) \ket{v_0}$, with $\ket{v_0}=\xi P\ket{\varphi}$.
    Thus, from \eqref{eq:sum-pup-ptupt}, \eqref{eq:ident-pup-with-sqrt-u-and-time-rev-op} and the definitions in \eqref{eq:gram-hadamard-test-implicit-first-state} we obtain
    \begin{subequations}
    \begin{align}
        \frac{1}{\xi^2}\left(\xi \bra{\varphi}P\right) U \left(\xi P\ket{\varphi}\right) +
        \left(1-\frac{1}{\xi^2}\right)
        \left(\xi^{\perp}\bra{\varphi} P^{\perp}\right) U \left(\xi^{\perp} P^{\perp} \ket{\varphi}\right)
        =&
        \frac{1}{\xi^2}\bra{v_0} U \ket{v_0} +
        \left(1-\frac{1}{\xi^2}\right)
        \bra{v_0^{\perp}} U \ket{v_0^{\perp}}\\
        =& \bra{\varphi}\hermpart(U)\ket{\varphi},
    \end{align}
    \end{subequations}
    with $\xi^{\perp}=\frac{\xi}{\sqrt{\xi^2-1}}$.
    The latter equation corresponds to \eqref{eq:gram-hadamard-test-implicit-proof-i},
    which together with \eqref{eq:re-innerp-equiv-herm-part} confirms the claim.
\end{proof}

We obtain the equivalent result to \autoref{thm:gram-hadamard-test-implicit} for the overlap with the Hamiltonian $\myham{}$.
\begin{theorem}
    \label{thm:H-overlap-hadamard-test-implicit}
    Under the same conditions and definitions as \autoref{thm:gram-hadamard-test-implicit}, we have that
    \begin{equation}
        \label{eq:H-overlap-hadamard-test-implicit}
        \boxed{
        \frac{\imath}{\xi^2} \bra{v(\htimedab)}(\imath\myham{}T)\ket{v(\htimedab)} -
        \imath\left(1-\frac{1}{\xi^2}\right) \bra{v^{\perp}(\htimedab)}(\imath\myham{}T)\ket{v^{\perp}(\htimedab)} =
        \imath \impart \bra{\varphi}U\myham{}\ket{\varphi}.}
    \end{equation}
\end{theorem}
\begin{proof}
    The pattern for this proof is very similar to that of \autoref{thm:gram-hadamard-test-implicit}.
    For a complex scalar $z\in \mathbb{C}$ we denote by $\impart z\coloneqq(z-\overline{z})/(\imath 2)$ its imaginary part.
    First, by the commutativity of $\myham{}$ and $U^{\dagger}$ ($U$ defined in \autoref{thm:gram-hadamard-test-implicit}) we have that $\skewhpart(U\myham{})=\skewhpart(U)\myham{}$, then
    \begin{align}
        \label{eq:im-innerp-equiv-herm-part}
        \imath\impart \bra{\varphi}U\myham{}\ket{\varphi}=\bra{\varphi}\skewhpart(U)\myham{}\ket{\varphi}.
    \end{align} 
    We aim to prove the following \begin{align}
        \label{eq:H-overlap-hadamard-test-implicit-proof-i}
        \frac{1}{\xi^2} \bra{v(t_a)}\myham{}\ket{v(t_b)} +
        \left(1-\frac{1}{\xi^2}\right) \bra{v^{\perp}(t_a)}\myham{}\ket{v^{\perp}(t_b)} =&
        \bra{\varphi}\skewhpart(U)\myham{}\ket{\varphi},
    \end{align}
    which, as a result of \autoref{lemma:overlap-with-K-from-expectation-on-KT} (with $K=\myham{}$),
    is equivalent to the claim in \eqref{eq:H-overlap-hadamard-test-implicit}.

    Using the identities in \autoref{lemma:proj-identities} we obtain
    \begin{subequations}
    \begin{align}
        \label{eq:puhp_plus_ppuhpp_eq_skewuh}
        P U \myham{} P + P^{\perp} U \myham{} P^{\perp} \underset{\eqref{eq:ident-proj-hamilt}}{=}&
        (P U P^{\perp})\myham{} + (P^{\perp} U P)\myham{}\\
        \underset{\eqref{eq:ident-proj-skewhp-time-evol}}{=}&
        \skewhpart(U) P^{\perp} \myham{} + \skewhpart(U) P \myham{}\\
        =& \skewhpart(U) \myham{},
    \end{align}
    \end{subequations}
    where the last equality follows from the complementarity of the projectors.

    Starting from \eqref{eq:puhp_plus_ppuhpp_eq_skewuh}, that is $P U \myham{} P + P^{\perp} U \myham{} P^{\perp}=\skewhpart(U) \myham{}$, we get
    \begin{align}
        \frac{1}{\xi^2}\left(\xi \bra{\varphi}P\right) U \myham{} \left(\xi P\ket{\varphi}\right) +
        \left(1-\frac{1}{\xi^2}\right)
        \left(\xi^{\perp}\bra{\varphi} P^{\perp}\right) U \myham{} \left(\xi^{\perp} P^{\perp} \ket{\varphi}\right)
        =\bra{\varphi}\skewhpart(U)\myham{}\ket{\varphi},
    \end{align}
    with $\xi^{\perp}=\frac{\xi}{\sqrt{\xi^2-1}}$.
    Now, since $U=\exp\left(\imath t_a\myham{}\right) \exp\left(-\imath t_b\myham{}\right)$ and since each factor commutes with $\myham{}$,
    we rearrange the LHS of the preceding equation to obtain \eqref{eq:H-overlap-hadamard-test-implicit-proof-i}, which in turn confirms the claim.
\end{proof}

\begin{figure}
    \centering
    \includegraphics[width=\linewidth]{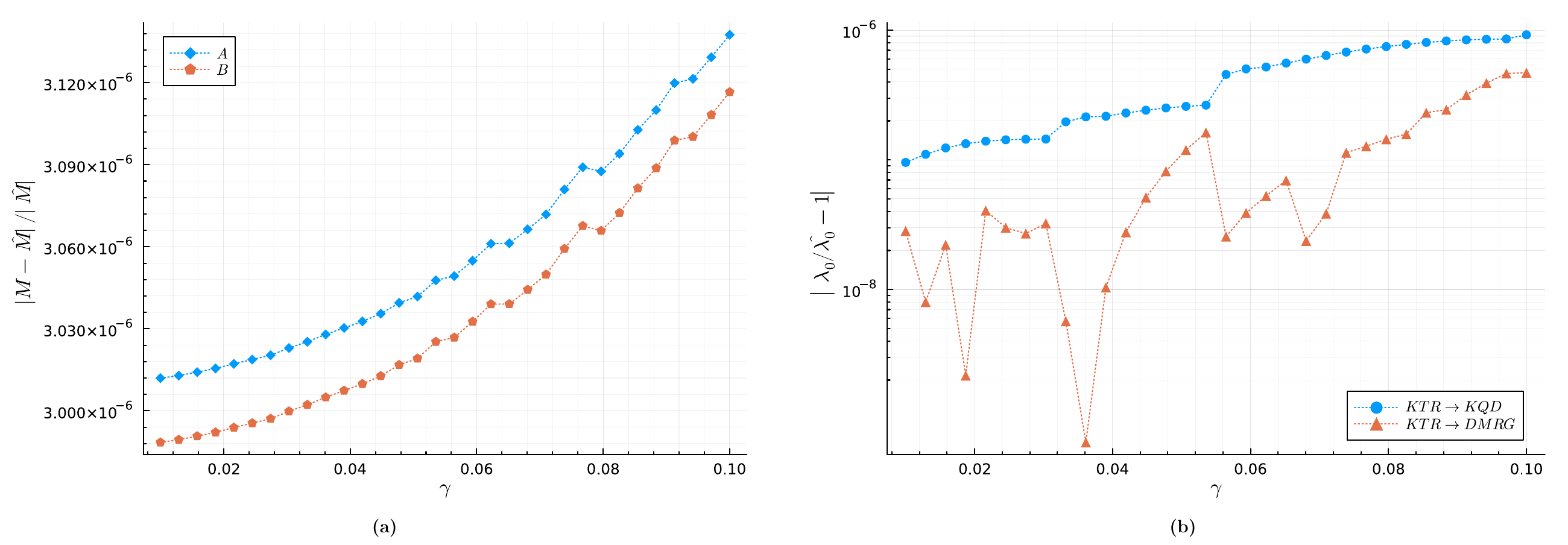}
    \caption{
    Numerical demonstration (MPS) of \autoref{thm:gram-hadamard-test-implicit}
    and \autoref{thm:H-overlap-hadamard-test-implicit} (implicit Hadamard test)
    for the TFIM on 64 qubits, reflecting the case of \autoref{section:example-ising} and \autoref{fig:experiment-tfim}.
    \textbf{(a)}
    Relative errors for the overlap matrices $A$ and $B$,
    where reference quantities are computed using the RHSs of \eqref{eq:gram-hadamard-test-implicit} and \eqref{eq:H-overlap-hadamard-test-implicit}.
    \textbf{(b)}
    Relative error of the ground energy $\lambda_0$ computed using the overlaps on the LHSs of \eqref{eq:gram-hadamard-test-implicit} and \eqref{eq:H-overlap-hadamard-test-implicit}.
    As references, we consider the canonical KQD ($\bullet$) and DMRG ($\blacktriangle$).
    }
    \label{fig:implicit-Hadamard-Ising}
\end{figure}
In practice, the foregoing results determine the overlap matrices whose entries are defined by
\begin{subequations}
\begin{align}
    \label{eq:overlaps-from-implicit-hadamard}
    A_{a, b} =&
    \imath \impart \bra{\varphi}U(t_b-t_a) \myham{}\ket{\varphi},\\
    B_{a, b} =& \repart \bra{\varphi}U(t_b-t_a) \ket{\varphi},
\end{align}
\end{subequations}
with $U(t)=\exp\left(-\imath t\myham{}\right)$.
The computations of the imaginary and real parts, are obtained using the LHS of respectively \eqref{eq:H-overlap-hadamard-test-implicit} and \eqref{eq:gram-hadamard-test-implicit}.
Put differently, as a result of \eqref{eq:skew-and-herm-parts-to-sin-cos} and \eqref{eq:re-innerp-equiv-herm-part},
this could be interpreted as preparing a Krylov subspace spanned by the time evolution of the cosine.
In \autoref{fig:implicit-Hadamard-Ising} we provide a numerical demonstration of the implicit Hadamard test, using the same model considered in \autoref{section:example-ising}.
The relative error for the ground energy is comparable with the previous results depicted in \autoref{fig:experiment-tfim} (specifically, see the series corresponding to $s=2$).

\section{Localized projection circuit for initial states}
\label{section:proj-circ}
We illustrate a strategy related to the construction of the initial states $\ket{v_0}, \ket{v_0^{\perp}}$, starting from an arbitrary state $\ket{\varphi}$, which we introduced in \autoref{section:init-state}.
In particular, we focus on the near-term implementability, so we consider states with \underline{localized entanglement}.
In essence, this formalizes the blocking pattern we used in the experiments in \autoref{section:example-ising} and \autoref{section:example-lgt}.
To achieve such a purpose, instead of considering the two complementary projectors $P$ and $P^{\perp}$ defined in \eqref{eq:projectors-def},
we construct a set of projectors $\{P_i\}$ characterized by a convenient entanglement structure.
We show that such a set can be partitioned into two subsets $\mathcal{P} \cup \mathcal{P}^{\perp}$, each distinguished by its elements $P_i$ being orthogonal to either $P$ or $P^{\perp}$.

We assume that the time reversal operator $T$ is tensor separable into $s$ operators\footnote{
    This condition is non-restrictive since for all models under consideration, the operator $T$
    consists of a tensor product of Pauli operators.
    In general, this is expected to hold for $k$-local Hamiltonians.
}, that is $T=\bigotimes_{i=1}^s T_b$.
Since $T$ is Hermitian-involutory, the same property applies to the block operator $T_b$.
In addition, we assume that the number of blocks $s \ll n$, where $n$ is the number of qubits\footnote{
    Moreover, the number of blocks $s$ is expected to depend at least linearly on the number of qubits $n$.
    This prevents the overlap between the ground state and the initial state from decaying exponentially as $n$ increases.
}.
Consider the functions $\alpha_i: \{1, \ldots, s\} \to \{0, 1\}$, such that $i_1\ne i_2$ implies $\sum_j |\alpha_{i_1}(j)-\alpha_{i_2}(j)|\ne 0$.
There are $2^s$ distinct functions $\alpha_i$.
Define the $i$th projector by
\begin{align}
    \label{eq:localized-ith-proj-struct}
    P_i \coloneqq& \bigotimes_{j=1}^s\frac{\idenmnodim + (-1)^{\alpha_i(j)}T_b}{2},
\end{align}
which is indeed an orthogonal projector since $P_i^2=P_i$ and $P_i^{\dagger}=P_i$.
The defining property of the functions $\alpha_i$ implies that $P_{i_1}P_{i_2}=0$ whenever $i_1\ne i_2$.
Let $p$ be the mapping on the set of functions ${\alpha_i}$, with rule $\alpha_i \mapsto (-1)^{\sum_j\alpha_i(j)}$.
Then we partition the set of projectors $\{P_i\}$ into two subsets conditioned to value of the function\footnote{
    We note that the function $p$ can be interpreted as a parity function, since the sign of its image depends on the parity of $\sum_j \alpha_i(j)$.
} $p$, so
\begin{align}
    \label{eq:local-entanglement-projectors}
    \mathcal{P} \coloneqq \{P_i|p(\alpha_i)=1\}, &\quad
    \mathcal{P}^{\perp} \coloneqq \{P_i|p(\alpha_i)=-1\}.
\end{align}
It can be shown that $TP_i=p(\alpha_i) P_i$, then
\begin{align}
    \label{eq:local-proj-mul-big-proj}
    PP_i = P_i \frac{1+p(\alpha_i)}{2},&\quad
    P^{\perp}P_i = P_i \frac{1-p(\alpha_i)}{2},
\end{align}
for all $i$, so we have either $PP_i = P_i$ or $P^{\perp}P_i =P_i$.
Hence, the sets $\mathcal{P}$ and $\mathcal{P}^{\perp}$ in \eqref{eq:local-entanglement-projectors} can be defined equivalently as
\begin{align}
    \label{eq:local-proj-subsets}
    \mathcal{P}=\{P_i|PP_i \ne 0\}, &\quad
    \mathcal{P}^{\perp}=\{P_i|P^{\perp}P_i\ne 0\}.
\end{align}

For all $i$, and any (conformable) state $\ket{\varphi}$, we observe the following property
\begin{align}
    \left(T - p(\alpha_i) \idenmnodim\right) P_i \ket{\varphi}=& 0,
\end{align}
which corresponds to the stabilizer property in \eqref{eq:vzero-stab}, with $\ket{v_0}=\xi P_i \ket{\varphi}$ and $c=p(\alpha_i)$ (see also \autoref{section:init-state}),
for some normalization coefficient $\xi >0$ (assuming $P_i \ket{\varphi}\ne 0$).

\begin{lemma}
    Let $P$ and $P^{\perp}$ be the projectors defined in \eqref{eq:projectors-def} and $P_i$ those in \eqref{eq:localized-ith-proj-struct}, then
    \begin{align}
        \sum_{P_i \in \mathcal{P}} P_i=P,\quad \sum_{P_i \in \mathcal{P}^{\perp}} P_i=P^{\perp},
    \end{align}
    so $\sum_i P_i =\idenmnodim$.
\end{lemma}
\begin{proof}
    For a single block, we note that $(\idenmnodim + T_b)/2 + (\idenmnodim - T_b)/2 =\idenmnodim$, then consider the induction step where the projectors $P_i$ act on $s$ blocks.
    Assuming $\sum_i P_i =\idenmnodim$, then for $s+1$ blocks we have that
    \begin{align}
        \sum_i \left(\frac{\idenmnodim + T_b}{2} \otimes P_i + \frac{\idenmnodim - T_b}{2} \otimes P_i \right) =\idenmnodim \otimes \idenmnodim.
    \end{align}
    The proof is concluded by observing that 
    $\sum_i P_i =\idenmnodim\, \implies P\sum_i P_i =P
    \underset{\eqref{eq:local-proj-subsets}}{\implies} \sum_{P_i \in \mathcal{P}} P_i = P$,
    and likewise for the second claim.
\end{proof}
This proves that the set $\{P_i\}$ is a complete orthogonal set of projectors.

We now have the required structures for extending \autoref{thm:gram-hadamard-test-implicit} as follows.
\begin{theorem}
    \label{thm:gram-hadamard-test-implicit-extended}
    For any pair of time displacements $t_a, t_b \in \mathbb{R}$, we have that
    \begin{align}
        \label{eq:gram-hadamard-test-implicit-extended}
        \sum_{i=1}^{2^s} p(\alpha_i) \bra{\varphi}P_i \left(\sqrt{U}^{\dagger} T \sqrt{U}\right) P_i\ket{\varphi} =&
        \repart
        \bra{\varphi}  \sum_{i=1}^{2^s}\left(P_i U P_i\right) \ket{\varphi},
    \end{align}
    with $U=\exp\left(-\imath (t_b-t_a)\myham{}\right)$.
    Also $\sqrt{U}=\exp\left(-\imath h\myham{}\right)$ with $h=(t_b-t_a)/2$.
\end{theorem}
In particular, when $\{P_i\}=\{P, P^{\perp}\}$, then as a consequence of \autoref{lemma:proj-identities}, the preceding theorem reduces to \autoref{thm:gram-hadamard-test-implicit}.
An equivalent extension for \autoref{thm:H-overlap-hadamard-test-implicit} can be obtained using the same reasoning; we omit the result.
\begin{proof}
    The proof follows the same reasoning as in \autoref{thm:gram-hadamard-test-implicit}. We outline the key steps.
    From \eqref{eq:local-proj-mul-big-proj} we have that either $P_iP\ne 0$ or $P_iP^{\perp}\ne 0$, assuming the former we have that $P_iP= P_i$.
    So from \eqref{eq:ident-proj-hermp-time-evol} and \eqref{eq:ident-pup-with-sqrt-u-and-time-rev-op} it follows that
    \begin{subequations}
    \begin{align}
        P_i U P_i =& P_i \hermpart(U) P_i,\\
        P_i U P_i =& p(\alpha_i) P_i\sqrt{U}^{\dagger}T\sqrt{U}P_i.
    \end{align}
    \end{subequations}
    When $P_iP^{\perp}\ne 0$, the result follows similarly, with attention to the negative sign in \eqref{eq:ident-pup-with-sqrt-u-and-time-rev-op}, which is realized by the
    expression $p(\alpha_i)$.
    We note that the mapping $U \mapsto \sum_i P_i U P_i$ is a projection.
    Hence, the above identities and the relation in \eqref{eq:re-innerp-equiv-herm-part},
    establish the claim in \eqref{eq:gram-hadamard-test-implicit-extended}.
\end{proof}

\begin{figure}
    \centering
    \begin{equation*}
    \vcenter{
    \Qcircuit @C=0.7em @R=.5em {
    \lstick{\ket{0}}   & \qw      & \qw      & \qw                  & \qw        & \gate{H}  & \ctrl{1}  & \gate{H}  & \qw       & \meter     & \rstick{i_1} \cw \\
    \lstick{}          & \qw      & \qw      & \gate{U_1^{\dagger}} & \qw        & \ctrl{1}  & \targ     & \ctrl{1}  & \qw       & \gate{U_1} & \qw \\
    \lstick{}          & \qw      & \qw      & \gate{U_2^{\dagger}} & \ctrl{1}   & \targ     & \qw       & \targ     & \ctrl{1}  & \gate{U_2} & \qw \\
    \lstick{}          & \qw      & \qw      & \gate{U_3^{\dagger}} & \targ      & \qw       & \qw       & \qw       & \targ     & \gate{U_3} & \qw \\
    \lstick{}          & \qw      & \qw      & \gate{U_1^{\dagger}} & \targ      & \qw       & \qw       & \qw       & \targ     & \gate{U_1} & \qw \\
    \lstick{}          & \qw      & \qw      & \gate{U_2^{\dagger}} & \ctrl{-1}  & \targ     & \qw       & \targ     & \ctrl{-1} & \gate{U_2} & \qw \\
    \lstick{}          & \qw      & \qw      & \gate{U_3^{\dagger}} & \qw        & \ctrl{-1} & \targ     & \ctrl{-1} & \qw       & \gate{U_3} & \qw
    \inputgroupv{2}{7}{.8em}{5.6em}{\ket{\varphi}}
    \gategroup{2}{4}{4}{10}{.5em}{.}
    \gategroup{2}{11}{7}{11}{.8em}{\}}
    \\
    \lstick{\ket{0}}   & \qw     & \qw      & \qw                  & \qw        & \gate{H}  & \ctrl{-1} & \gate{H}  & \qw       & \meter     & \rstick{i_2} \cw
    }} \quad = \quad
    \xi_i
    P_i\ket{\varphi}
    \end{equation*}
    \caption{
    General case for the projection $P_i$ applied to the arbitrary state $\ket{\varphi}$ (on $n=6$ qubits, with $s=2$ blocks) when the time reversal operator is of the form
    $T=\bigotimes_{l=1}^n V_l$, where $V_l$ is a non-identity, single-qubit involutory unitary (e.g. Paulis).
    This pattern can be extended to any number of blocks, with each block requiring an additional qubit.
    The unitaries $U_l$ satisfy the equation $U_l \pauliopx{} U_l^{\dagger}=V_l$, where $V_l$ is the target unitary (with eigenvalues $\{\pm 1\}$).
    The bits $i_1, i_2$ define the index $i=2 i_1 + i_2$ determining the projection $P_i$ in \eqref{eq:localized-ith-proj-struct}.
    The coefficients $\xi_i \in (0, +\infty]$ are defined such that $\sum_i 1/\xi_i^2=1$.
    We note that the state is defined when $\xi_i$ is finite, that is, the probability $1/\xi_i^2$ is non-zero.
    }
    \label{fig:localized-proj-circuit}
\end{figure}
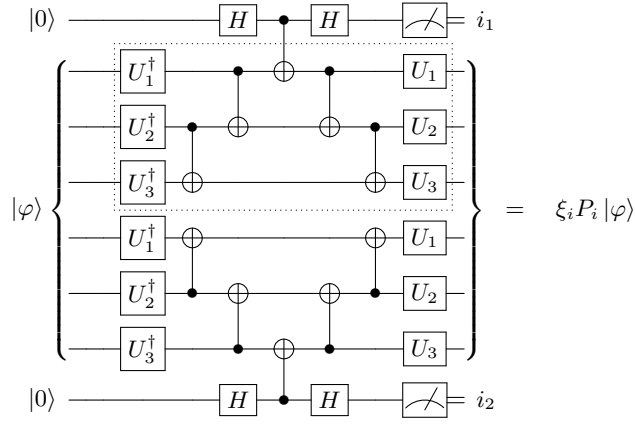

\begin{figure}
    \centering
    \includegraphics[width=\textwidth]{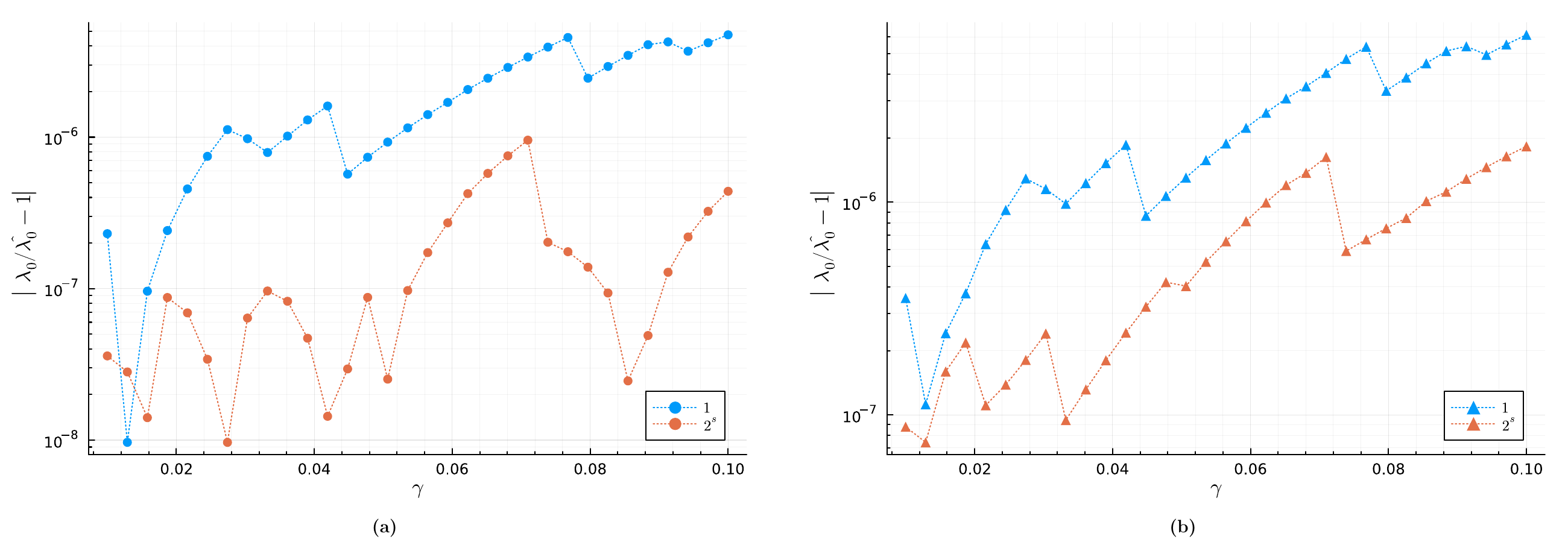}
    \caption{
        Numerical demonstration (MPS) of \autoref{thm:gram-hadamard-test-implicit-extended}
        for the TFIM on 64 qubits, reflecting the case of \autoref{section:example-ising} and \autoref{fig:experiment-tfim}, with $s=4$ blocks.
        We plot the relative error of the ground energy $\lambda_0$ computed using the initial states produced by the circuit in \autoref{fig:localized-proj-circuit},
        with $\ket{\varphi}=\ket{+}^{\otimes n}$.
        As references, we consider the canonical KQD ($\bullet$, plot (a)) and DMRG ($\blacktriangle$, plot (b)).
        Here, the series are distinguished by the number of terms in \eqref{eq:gram-hadamard-test-implicit-extended} (related to the projectors $P_i$).
        The case $2^s$ (i.e., all projectors) shows improved convergence.
    }
    \label{fig:experiment-implicit_hadamard_extended}
\end{figure}

Using the following circuit identity recursively
\begin{equation}
    \vcenter{
    \Qcircuit @C=0.7em @R=.5em {
         & \ctrl{1} & \ctrl{2} & \qw \\
         & \targ & \qw & \qw \\
         & \qw & \targ & \qw
    }} \,=\,
    \vcenter{
    \Qcircuit @C=0.7em @R=.5em {
         & \qw & \ctrl{1} & \qw & \qw \\
         & \ctrl{1} & \targ & \ctrl{1} &\qw \\
         & \targ & \qw & \targ & \qw
    }}
\end{equation}
in \autoref{fig:localized-proj-circuit} we devise a circuit construction for the preparation of the initial states $\{P_i \ket{\varphi}\}$, based on the projectors in \eqref{eq:localized-ith-proj-struct}.
Such states can be used in conjunction with \autoref{thm:gram-hadamard-test-implicit-extended}, to obtain the overlap matrices following the approach taken in \eqref{eq:overlaps-from-implicit-hadamard}.
In \autoref{fig:experiment-implicit_hadamard_extended} we provide a numerical demonstration of \autoref{thm:gram-hadamard-test-implicit-extended}, using the same model considered in \autoref{section:example-ising}.

\begin{remark}
    In relation to the circuit in \autoref{fig:localized-proj-circuit}, when $(\bigotimes_{j} U_j^{\dagger})^{\otimes s}\ket{\varphi}$ is a computational basis state or a
    linear combination of fixed points of the permutation implemented by the first ladder of CNOTs, then the structure simplifies to a product of GHZ-like states\footnote{
        We call these states GHZ-like since the circuit mirrors the structure of the 
        \textit{Greenberger–Horne–Zeilinger} state preparation.
    } (without requiring additional qubits).
    The latter is the case for the benchmarks we considered in \autoref{section:example-ising} and \autoref{section:example-lgt}.
\end{remark}

\section{Additional Theoretical Considerations}
\label{section:more-theory}
\subsection{Overlap by differentiation and integration}
In \autoref{section:quantum-krylov-method}, we have shown that the KQD method requires the computation of the overlap matrices $A$ and $B$,
defining the generalized eigenvalue problem.
However, when the conditions of \autoref{section:formulation} are fulfilled, we prove that we can obtain the entries of matrix $B$ from the entries of $A$ and vice versa, using quadrature \cite{Golub1967CalculationOG} or derivative discretization \cite{byrne-superkrylov} methods.

The result that follows can be framed as a further consequence of the implication $\{\myham{}, T\}=0\,\implies \frac{1}{2}[\imath \myham{}, T]=\imath\myham{}T$.
Before, we introduce the \textit{Kubo formula} \cite{doi:10.1143/JPSJ.12.570, 10.1063/1.526596}. For square matrices $A, B$ of the same size, we have that
\begin{align}
    \label{eq:kubo}
    \left[A, e^{-\imath t B}\right] =& -\imath e^{-\imath t B} \int_{0}^{t} e^{\imath \tau B} [A, B] e^{-\imath \tau B}\,\mathrm{d}\tau,
\end{align}
for all $t\in \mathbb{R}$.
\begin{lemma}  
    \label{lemma:integral-derivative-relation}
    Assuming the conditions of \autoref{lemma:gram-from-expectation-on-T} and \autoref{lemma:overlap-with-K-from-expectation-on-KT} (with $K=\myham{}$), we have that
    \begin{subequations}
    \begin{align}
        \label{eq:integrate-H-overlap}
        2c \int_0^{\htimedab} \bra{v(\tau)} (\imath \myham{} T) \ket{v(\tau)}\,\mathrm{d}\tau + 1 =& c\bra{v(\htimedab)}T\ket{v(\htimedab)},\\
        \label{eq:dt-gram}
        \frac{\imath c}{2}\frac{\mathrm{d}}{\mathrm{d}\htimedab} \bra{v(\htimedab)}T\ket{v(\htimedab)} =& \imath c\bra{v(\htimedab)}(\imath \myham{}T)\ket{v(\htimedab)}.
    \end{align}
    \end{subequations}
\end{lemma}
\begin{proof}
    Considering the Hamiltonian $\myham{}$ and a corresponding time reversal operator $T$,
    we rewrite (substituting $A \leftarrow T$ and $B \leftarrow \myham{}$) the Kubo formula in \eqref{eq:kubo} as follows
    \begin{align}
        \int_0^{\htimedab} e^{\imath \tau \myham{}} [\imath \myham{}, T] e^{-\imath \tau \myham{}}\,\mathrm{d}\tau =& 
        e^{\imath \htimedab{}\myham{}} [T, e^{-\imath \htimedab{} \myham{}}],
    \end{align}
    where the RHS sandwiched between $\bra{v_0}\cdots\ket{v_0}$ expands as
    \begin{subequations}
    \begin{align}
        \bra{v_0}
        e^{\imath \htimedab{}\myham{}} [T, e^{-\imath \htimedab{} \myham{}}]
        \ket{v_0} =&
        \bra{v_0} e^{\imath \htimedab{}\myham{}}Te^{-\imath \htimedab{} \myham{}} \ket{v_0}
        - \bra{v_0} T \ket{v_0}\\
        \underset{\eqref{eq:v-at-t-def}, \eqref{eq:vzero-stab}}{=}&
        \bra{v(\htimedab)} T \ket{v(\htimedab)} - c.
    \end{align}
    \end{subequations}
    Then, noting that $c^2=1$ (as explained in \autoref{section:formulation}, $c\in \{\pm 1\}$)
    \begin{align}
        c\int_0^{\htimedab} \bra{v(\tau)} [\imath \myham{}, T] \ket{v(\tau)}\,\mathrm{d}\tau =& 
        c\bra{v(\htimedab)} T \ket{v(\htimedab)} - 1.
    \end{align}
    Hence, the claim in \eqref{eq:integrate-H-overlap} follows from the relation $\frac{1}{2}[\imath \myham{}, T]=\imath\myham{}T$.
    Using the same relation, we deduce the second claim, so
    \begin{subequations}
    \begin{align}
        \frac{1}{2}
        \frac{\mathrm{d}}{\mathrm{d}\htimedab} \bra{v(\htimedab)}T\ket{v(\htimedab)}
        \underset{\eqref{eq:v-at-t-def}}{=}&
        \frac{1}{2}
        \bra{v(\htimedab)}(\imath \myham{}T)\ket{v(\htimedab)} - 
        \frac{1}{2}
        \bra{v(\htimedab)}(\imath T\myham{})\ket{v(\htimedab)}\\
        =&
        \frac{1}{2} \bra{v(\htimedab)}[\imath \myham{},T]\ket{v(\htimedab)}\\
        =& \bra{v(\htimedab)}(\imath \myham{}T)\ket{v(\htimedab)}.
    \end{align}
    \end{subequations}
\end{proof}
In the preceding claim, we note that the arguments of the integral and derivative,
and the expressions on the RHSs, correspond interchangeably to the RHSs of main results in \eqref{eq:second-key-result} and \eqref{eq:first-key-result}
(up to a constant factor). 

\begin{figure}
    \centering
    \includegraphics[width=\linewidth]{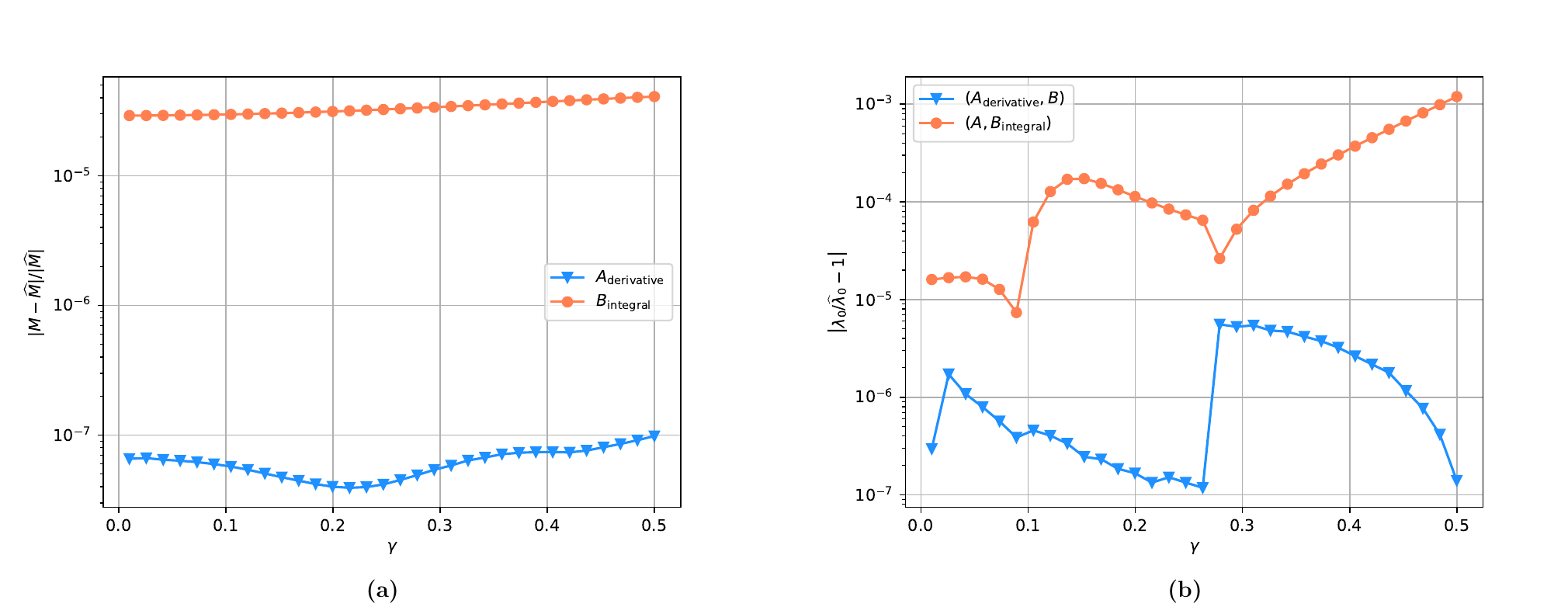}
    \caption{Numerical demonstration (MPS) of \autoref{lemma:integral-derivative-relation} for the TFIM on 48 qubits, with Krylov dimension $m=10$. The matrices $A_{\text{derivative}}$ and $B_{\text{integral}}$ are estimated numerically via the relations \eqref{eq:dt-gram} and \eqref{eq:integrate-H-overlap} respectively, using $20m$ sampling points. \textbf{(a)} Relative error for the overlap matrices $A_{\text{derivative}}$ and $B_{\text{integral}}$, with references to the quantities $A$ and $B$ computed via \eqref{eq:mat_a_entry-from-obs-iht} and \eqref{eq:first-key-result} respectively. \textbf{(b)} Relative error of the ground energy $\lambda_0$ of the pencils $(A_{\text{derivative}},B)$ and $(A,B_{\text{integral}})$, with reference to the ground energy of $(A,B)$.}
    \label{fig:implicit-Hadamard-Ising-estimates}
\end{figure}

Suppose $\myham{}=\sum_{i=1}^L h_i \boldsymbol{\sigma}_i$ for $\boldsymbol{\sigma}_i \in \{\idenmnodim, \pauliopx{}, \pauliopy{}, \pauliopz{}\}^{\otimes n}$ and $h_i\in\mathbb{R}$.
Computing the matrices $A$ and $B$ directly via \eqref{eq:mat_a_entry-from-obs-iht} and \eqref{eq:first-key-result} requires $m$ and $Lm$ expectation values, respectively.
A natural strategy, therefore, is to construct $B$ directly using \eqref{eq:first-key-result}, and then numerically estimate $A$ via the relation \eqref{eq:dt-gram}, with the inherent overhead of additional time samples required for accurate derivative estimation.
\autoref{fig:implicit-Hadamard-Ising-estimates} presents a numerical demonstration of this strategy, alongside the complementary approach where $A$ is computed directly using \eqref{eq:mat_a_entry-from-obs-iht}, and $B$ is estimated numerically via \eqref{eq:integrate-H-overlap}.
For derivative estimation, we employ the regularized state-space approach proposed in \cite{byrne-superkrylov}. For integral estimation, we utilize the composite Simpson's rule \cite{davis2007methods}.

\subsection{On proper Hermitians with time reversal symmetry}
The experiments we presented in \autoref{section:experiments}, concern exclusively (real) symmetric Hamiltonians,
so we check on the existence of proper Hermitians fulfilling the condition for the time reversal.

Let $K$ be a non-zero Hermitian such that $K^{\top}\ne K$ and $K^{\dagger}=K$.
Let $G$ be any conformable involutory Hermitian such that $[K, G]=0$ (possibly $G=\idenmnodim$).
Then $\myham{}=\pauliop{x}\otimes K$ is Hermitian and non-symmetric\footnote{
    Indeed, $\myham{}^{\top}=\pauliop{x}\otimes K^{\top}$, and by assumption $K^{\top}\ne K$, then
    $\myham{}^{\top}-\myham{}=\pauliop{x}\otimes (K^{\top}-K)\ne 0$.
    So it must be that $\myham{}^{\dagger}= \myham{}$.
}.
Let $T=\pauliop{z}\otimes G$, then $T^2=\idenmnodim$ and $T\myham{}T=(\pauliop{z}\pauliop{x}\pauliop{z})\otimes (GKG)=-\pauliop{x}\otimes K=-\myham{}$, that is $\{\myham{}, T\}=0$.
Hence, the existence claim is confirmed.

\clearpage
\pagebreak
\newpage

\bibliography{refs}

\end{document}